# QUALITY OF OCR FOR DEGRADED TEXT IMAGES

Roger T. Hartley, Kathleen Crumpton

Department of Computer Science, New Mexico State University

## ABSTRACT

Commercial OCR packages work best with high-quality scanned images. They often produce poor results when the image is degraded, either because the original itself was poor quality, or because of excessive photocopying. The ability to predict the word failure rate of OCR from a statistical analysis of the image can help in making decisions in the trade-off between the success rate of OCR and the cost of human correction of errors.

This paper describes an investigation of OCR of degraded text images using a standard OCR engine (Adobe Capture). The documents were selected from those in the archive at Los Alamos National Laboratory. By introducing noise in a controlled manner into perfect documents, we show how the quality of OCR can be predicted from the nature of the noise. The preliminary results show that a simple noise model can give good prediction of the number of OCR errors.

## INTRODUCTION

OCR (Optical Character Recognition) of scanned images of text documents is now generally considered to be a solved problem. Many commercial packages are available at low cost which do an impressive job on high quality originals. Using a variety of techniques culled from the research in the area, OmniPage (Caere), TextBridge (Xerox) and Capture (Adobe), among many others can give character recognition percentage rates in the high nineties. However, several things can conspire against attaining this rate consistently. In this paper we will ignore scanning issues, skew correction, text and paper color, and many other aspects of using an OCR package. Instead we concentrate on the quality of the original.

This can be a problem for the following reasons:

1. the original is old, and has suffered physical degradation.
2. the original was produced on a manual typewriter, so the individual characters can show variations in pressure and position.
3. the original is a carbon copy produced on a typewriter.
4. the original is a low quality photocopy, and shows variations in toner density and character spread, as well as general copier "grunge" caused by a dirty glass or background.

Any of these factors can contribute to less than acceptable OCR. The question is, how low can accuracy be allowed to go before the process becomes unable to produce usable online documents?

## THE ECONOMICS OF OCR

The aim of OCR is to turn hard copy text into a soft copy version for use online. The question of acceptable error rates has not been studied much, but a few observations based on common sense are useful in lieu of genuine results:

1. Fixing OCR errors is expensive since a human operator must read every word in both the original and the OCR version. False positives (words incorrectly identified by OCR) are rare but common enough to cause a problem. False negatives (words readable by the operator, but not identified by OCR) are much more common. The operator can also make errors in attempting the fix.
2. Errors that remain can cause frustration in users of the document, especially when the errors are in key words, or headings.

3. A technique, like that employed in Adobe Capture, where the image of the word is substituted, *in place,* when OCR fails, can alleviate some of this frustration. The downside to this is an inability to put these words (since they are images, not text) into indices or to use them in search.
4. The error rates for degraded documents can be so high as to make it increasingly unlikely that the end result is acceptable, even after manual editing.

In the light of this, we are left with two choices, achieved an impressive accuracy, but at a cost of 90 minutes per page. As an extreme solution, [Lesk] even suggests rekeying the document as an alternative.

The second choice is to budget for a certain amount of manual fix-up from a standard OCR package, but begs the question of how much is economically feasible.

## THE BASIC TECHNIQUE: CONTROLLED DEGRADATION

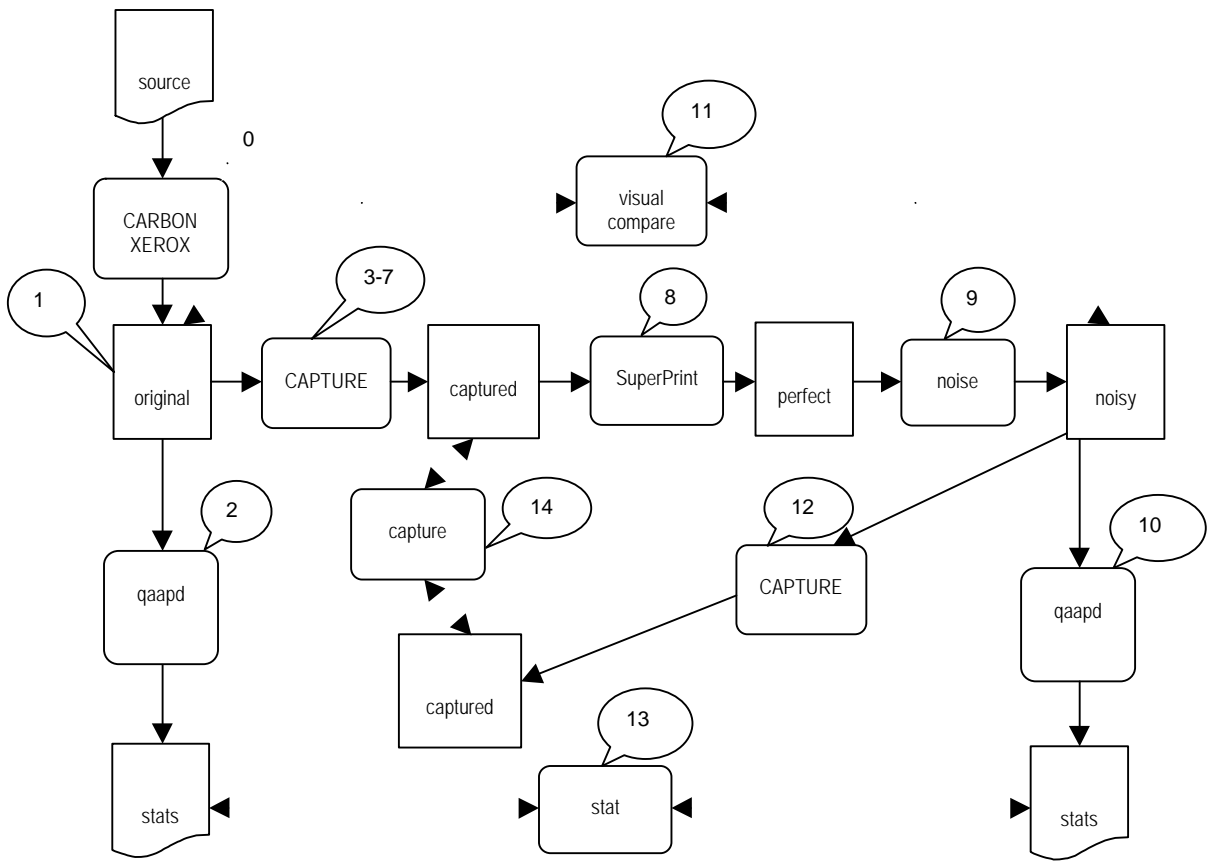

**Figure 1 The Degradation Procedure**

neither of which can be made comfortably. The first is to abandon standard OCR and improve it by using better techniques. Developing an OCR engine is a very difficult task, and in any case, there is a trade-off between OCR error rate and the cost (basically an issue of pages per minute) of doing that OCR. A general rule is that 95% accuracy (for perfect documents) is easy to achieve, but that getting closer to 100% requires much more computing power and OCR is much slower. For instance, [Kopec, 94] used a technique based on hidden Markov models, and

In this paper, we attempt to get a quantitative handle on this question using a technique based on known degradation levels in perfect documents. In order to provide a built-in validation procedure for the degradation model, we employ the following procedure:
1. Take a document page from the archive.
2. Analyze the image for black and white clusters.
3. Capture it to produce Adobe's hybrid text + image format.

4. Record the false positives (words that Capture did not suspect, but got wrong) and the false negatives (words that Capture suspected as wrong, but got right).
5. Accept all the suspects. This produces an all-text version with errors.
6. Correct all the errors by hand, both false positives and false negatives.
7. Correct font and character problems.
8. Save the perfect text version as a bitmap image using the SuperPrint™ TIFF printer driver.
9. Apply degradation algorithms in varying amounts to produce a degraded image.
10. Analyze the degraded image for black and white clusters.
11. Compare the degraded image with the original visually for a "ball-park" check on the noise models used.
12. Use Capture to repeat steps 3 and 4.
13. Compare results of analyses in 2 and 10.
14. Compare the results of Capture in steps 3 and 10.

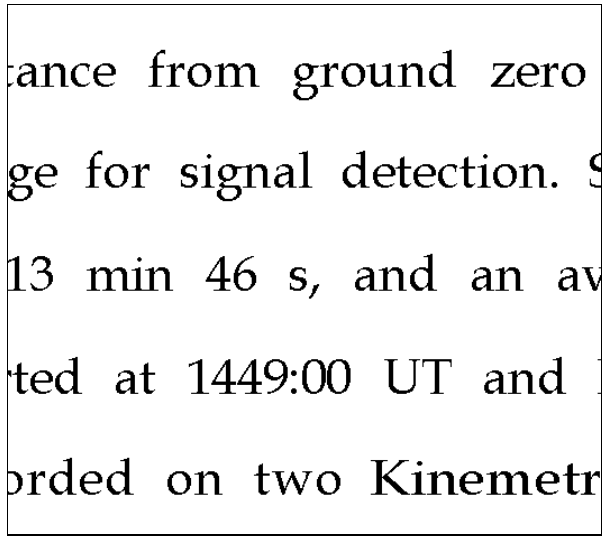

**Figure 2 A "perfect" document**

These steps are pictured in Figure 1, in which the square boxes represent data files, either bitmap images or output data, and the rounded boxes represent processes, either manual or computer. Qaapd is the program suite we wrote to perform the cluster analysis.

All images are uncompressed bi-level (black and white) files in 300 dpi TIFF format. We used three independently controlled techniques to produce an overall effect of degradation:
1. pixel noise: any number of black pixels placed at random anywhere on the page (x and y co-ordinates chosen independently from a flat distribution)
2. "blobs": any number of black or white pixels placed around a randomly chosen co-ordinate point and normally distributed about it; the spread of the normal distribution (essentially the standard deviation) is also variable. The white blobs can create broken characters whereas black blobs can create joined characters.
3. copy noise: each character (actually any connected set of black pixels) is grown by a small number of pixels (usually one or two) at random along the boundary of the character. This produces the familiar blurring of character boundaries as produced by a photocopier or scanner. Again a normal distribution was used.

Figures 2-5 show sections from example pages of these effects. The effects have been exaggerated so that it is clear what each degradation technique does, but they are easily controlled to produce any amount of degradation from very slight to making the document unreadable.

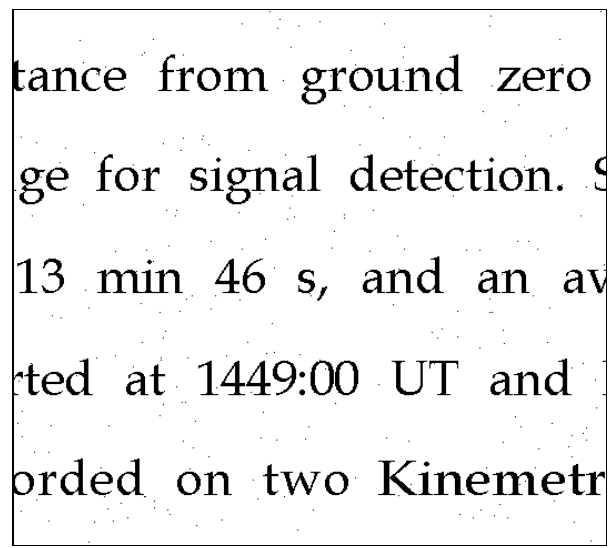

**Figure 3 With pixel noise added**

THE TEST DOCUMENT SET

Figures 8 and 9 shows extracts from documents in the archive at Los Alamos National Laboratory. From the sample it is clear that the image is less than perfect, although there is little evidence of noise caused by copying or general "grunge". However, OCR results are poor on documents like these. There can be many OCR errors, depending, it appears, mainly on the age of the original document. Many characters show breaks which presumably are faint areas on the original not picked up by the scanning process. On occasion there are "white lines" apparent (Figure 7) where much of a line of text is broken by a virtual white line. Presumably this was caused by a crease in the paper, or a segment of the typewriter ribbon that had been used before and had no ink left. The documents were selected to contain no graphics, tables or diagrams, but some contained mathematical formulae or equations. Some were in a fixed width Courier-like font, whereas some were in a Times-Roman-like font with serifs.

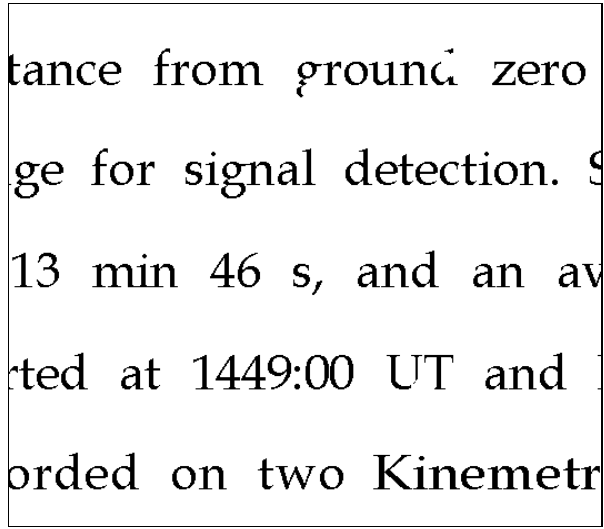

**Figure 4 With white "blobs" added**

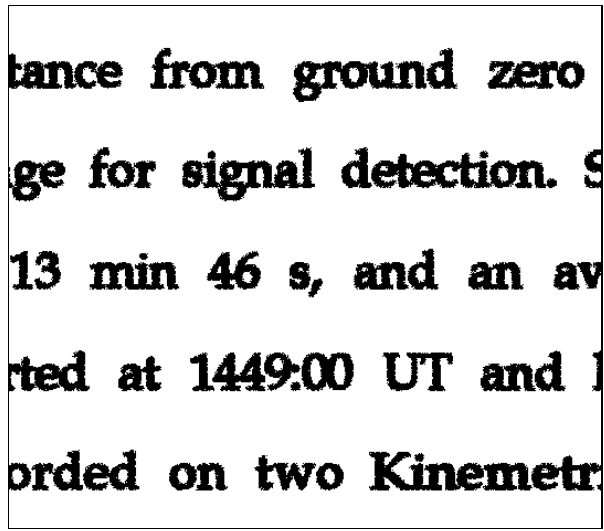

**Figure 5 With blurring added**

## DEGRADATION PARAMETERS

After many informal tests we used a combination of noise types 2 and 3 to form the degraded images. Type 3 degradation was applied first, using a 1 pixel blurring with a standard deviation of 4. Then white blobs were added using parameters of 0.1% of page pixels per blob (about 10000 pixels) and a spread of 6 pixels. This gave solid areas of white suitable for creating broken characters. 150 such blobs were added at random throughout the page. Many of course landed on white areas, so they had no effect, but between 20 and 40, on average, caused character breaks. Type 1 noise was not used since the chosen documents were mostly very clean, showing very little in the way of random single pixel noise.

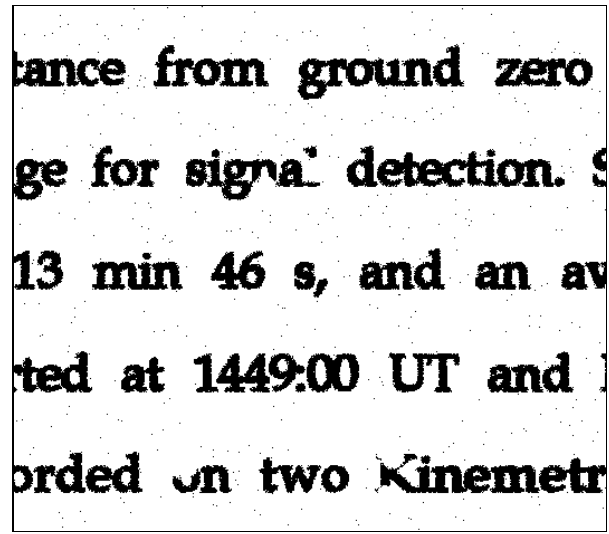

**Figure 6 With all three noise types added**

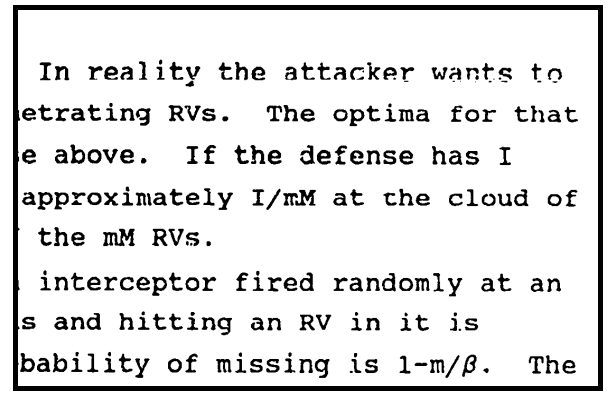

**Figure 7 An archive original showing the virtual white line**

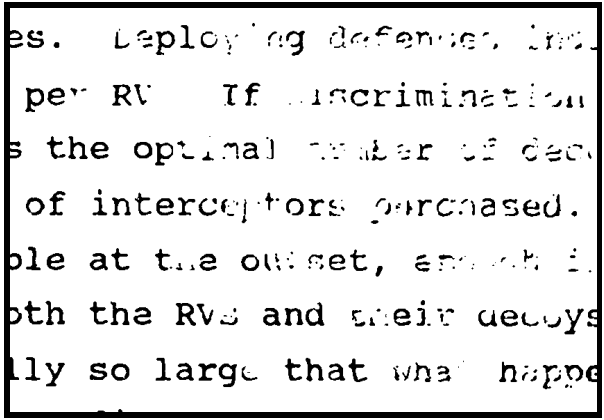

**Figure 8 A badly degraded document from the archive**

RESULTS: VISUAL COMPARISON

Figures 9 and 10 show a sample of an original document from the archive, and the same section from the same page processed by steps 1-9. They show a reasonable similarity, although clearly adjustment of the parameters could bring these two closer. However, the parameter settings that produced Figure 10 were found to give a reasonable spread across all the chosen documents.

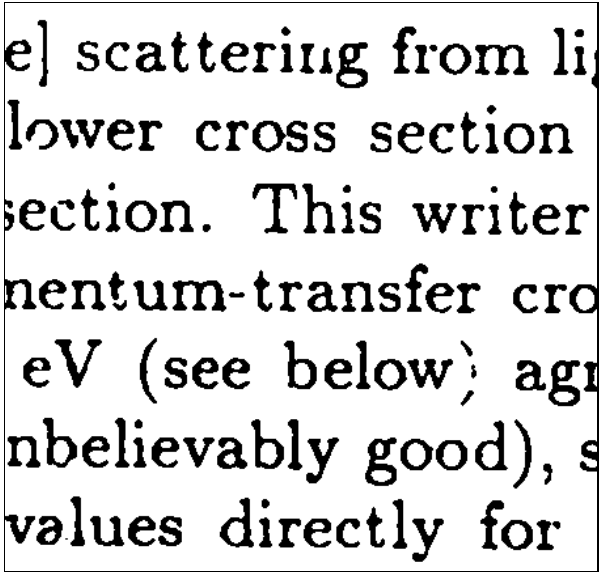

**Figure 9 A Section of an Original Document**

Capture is an example of an OCR engine that relies on word-level recognition. This means that individual characters are not flagged as errors, only whole words. Much of the work on noise models for document recognition has concentrated on character-level degradation [Jenkins et al., 94,

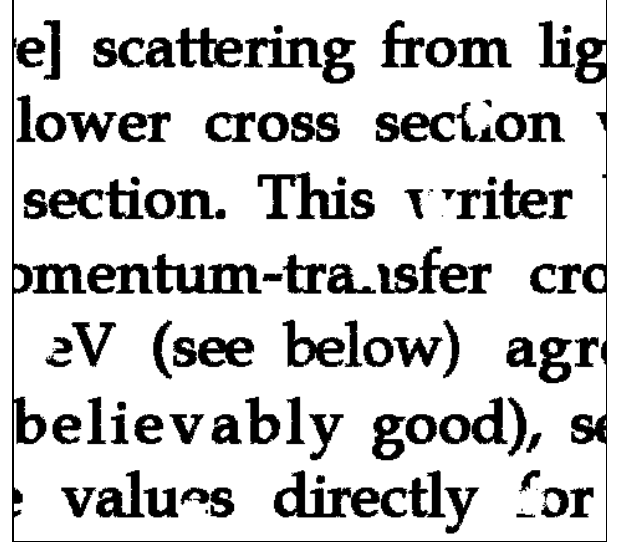

**Figure 10 The Noisy Equivalent to Figure 9**

RESULTS: CAPTURE

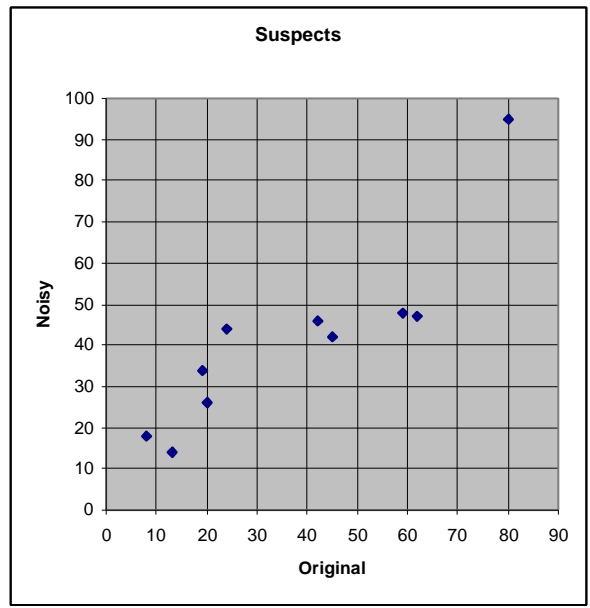

**Chart 1 Capture suspects**

Blando et al., 95, Cannon et al. 97], and thus is inappropriate for Capture. Preliminary results (discussed below) confirm this.

However, when the results of running Capture on the originals were compared with those obtained by running Capture on the noisy versions, a fairly close correspondence was obtained. Chart 1 summarises these results for ten selected pages used in the study. In the chart each point shows the number of suspects produced by Capture for the original document and the noisy version of it. (Again, a suspect is a whole word identified by Capture as having some recognition problem). Since Capture produces many false negatives (words it suspected, but actually got right) and a few false positives (words it thought it got right, but actually were wrong) we also made an attempt to capture the work needed to correct these types of error. The formula:

$$C = E + 0.7 F_+ + F_-$$

was used in another scatter plot, Chart 2, again showing a reasonable correlation.

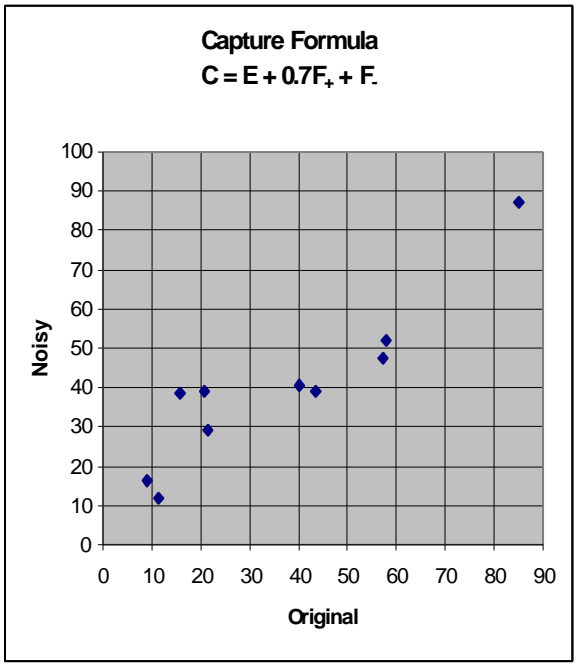

**Chart 2 Capture Estimated Work Formula**

RESULTS: STATISTICAL

As mentioned above, other researchers (cited above) used character-level degradation models in their work. The two common contributing elements in this work are the broken character problem, and the touching character problem. The first produces a higher proportion of black pixel clusters of small size, and the latter produces more black pixel clusters of large size. These clearly are features that can be the subject of a statistical analysis, such as in [Blando et al., 95], and indeed they obtained good results. An indication of this can be seen in Figures 11 and 12, which show a cluster analysis of an original and the perfect equivalent.

The histogram for the original shows clumps below a cluster size of 100 (broken characters) and again at 300 and above (touching characters) that are absent in the prefect version. We ran a linear regression on the set of originals using the two factors BSFH (black speckle factor-high, or clusters with size > 300) and BSFL (black speckle factor-low, or clusters with size <= 10) as independent variables, and the number of Capture suspects as the dependent variable. The resulting formula was then used to predict the suspects for the noisy documents generated from their perfect versions. The results are shown in Chart 3, which shows predicted suspects vs. actual suspects. Although there is a clear trend, the correlation is not very good. We can criticize the results for their small sample size (10), but our guess is that the noise model is not appropriate for word-level recognition engines like Capture. Further work is warranted to look for a better model. We plan to extend this small study to include more documents, and do a much more thorough analysis to uncover such a model.

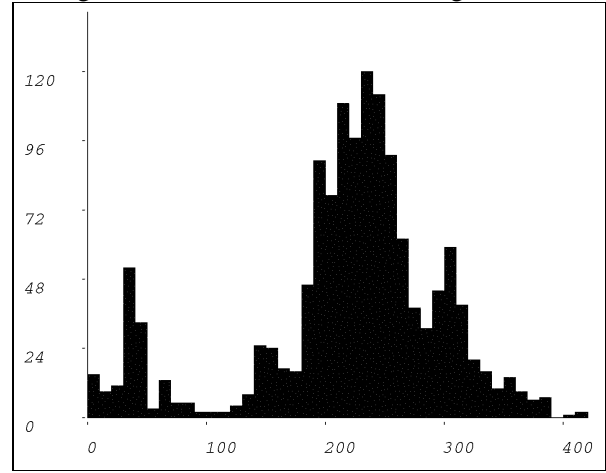

**Figure 11 Histogram of black cluster size vs. number of clusters for an original document.**

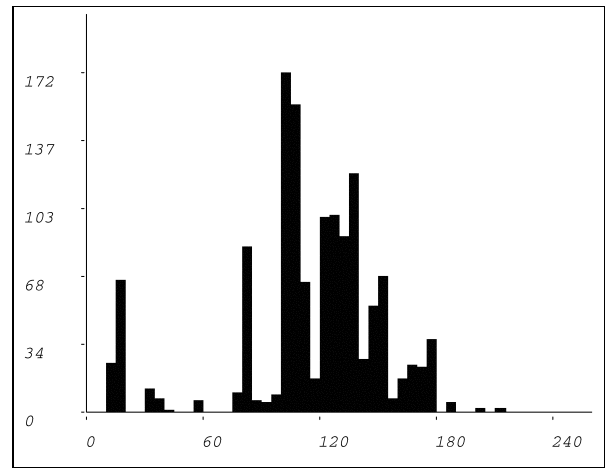

**Figure 12 Histogram of a perfect document corresponding to Figure 11.**

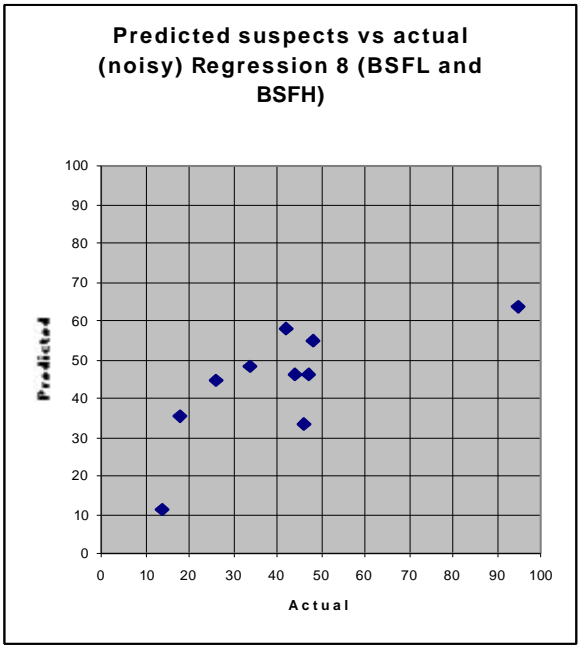

**Chart 3 Predicted suspects from regression analysis**

## CONCLUSION

Investigating the OCR of documents of poor quality is important for those whose digital libraries are being generated from archives of originals created before the digital age. We have discussed the use of one OCR engine, Capture, and discussed how the nature of the degradation can affect the accuracy of the resultant OCR effort. A character-level noise model was shown to give reasonably similar degraded versions, thus validating the general form of the model. However, the features used by other researchers as predictors of OCR accuracy were found to be insufficient for the word-level engine.

## ACKNOWLEDGEMENTS

The authors would like to acknowledge the help of Heather Pfeiffer for her programming skills, and Bob Webster of Los Alamos National Laboratory for supporting this work.